\documentclass[
 reprint,
 superscriptaddress,
 amsmath,
 amssymb,
 aps,
]{revtex4-2}

\usepackage{graphicx}
\usepackage{dcolumn}
\usepackage{bm}
\usepackage{siunitx}
\usepackage[dvipsnames]{xcolor}

\definecolor{PRXblue}{RGB}{0,115,180}

\usepackage[
    colorlinks=true,
    citecolor=PRXblue,
    linkcolor=PRXblue,
    urlcolor=PRXblue,
    hypertexnames=false
]{hyperref}

\DeclareSIUnit\angstrom{\text{Å}}
\DeclareSIUnit\gray{\text{Gy}}
\DeclareSIUnit[]{\photon}{ph}

\begin{document}

\preprint{APS/123-QED}

\title{On the Role of Atomic Mobility in the Polyamorphism of Metallic Glasses}

\author{Jie Shen}
\thanks{Contact author: \textnormal{jie.shen@unipd.it}}
\affiliation{Institut Néel, Université Grenoble Alpes and Centre National de la Recherche Scientifique, 25 rue des Martyrs - BP 166, 38042, Grenoble cedex 9, France}
\affiliation{European Synchrotron Radiation Facility, 71 avenue des Martyrs, CS 40220, Grenoble 38043, France}

\author{Antoine Cornet}
\affiliation{Institut Néel, Université Grenoble Alpes and Centre National de la Recherche Scientifique, 25 rue des Martyrs - BP 166, 38042, Grenoble cedex 9, France}
\affiliation{European Synchrotron Radiation Facility, 71 avenue des Martyrs, CS 40220, Grenoble 38043, France}

\author{Pei-Hao Sun}
\thanks{Contact author: \textnormal{peihao.sun@unipd.it}}
\affiliation{Department of Physics and Astronomy ``Galileo Galilei'', University of Padova, Via F. Marzolo, 8 - 35131 Padova, Italy}

\author{Valerio Di Lisio}
\affiliation{Donostia International Physics Center, Paseo Manuel de Lardizabal 4, 20018 San Sebastián, Spain}
\affiliation{Centro de Física de Materiales (CSIC–UPV/EHU), Paseo Manuel de Lardizabal 5, 20018 San Sebastián, Spain}

\author{Daniele Cangialosi}
\affiliation{Donostia International Physics Center, Paseo Manuel de Lardizabal 4, 20018 San Sebastián, Spain}
\affiliation{Centro de Física de Materiales (CSIC–UPV/EHU), Paseo Manuel de Lardizabal 5, 20018 San Sebastián, Spain}

\author{Klara Otto}
\affiliation{Chair of Metallic Materials, Technical University of Berlin, Ernst-Reuter-Platz 1, 10587 Berlin, Germany}

\author{Isabella Gallino}
\affiliation{Chair of Metallic Materials, Technical University of Berlin, Ernst-Reuter-Platz 1, 10587 Berlin, Germany}

\author{Martin Rosenthal}
\affiliation{European Synchrotron Radiation Facility, 71 avenue des Martyrs, CS 40220, Grenoble 38043, France}

\author{Eloi Pineda}
\affiliation{Department of physics, Institute of Energy Technologies, Center for Research in Multiscale Science and Engineering, Universitat Politècnica de Catalunya-BarcelonaTech, 08019 Barcelona, Spain}

\author{Giulio Monaco}
\affiliation{Department of Physics and Astronomy ``Galileo Galilei'', University of Padova, Via F. Marzolo, 8 - 35131 Padova, Italy}

\author{Marco Di Michiel}
\affiliation{European Synchrotron Radiation Facility, 71 avenue des Martyrs, CS 40220, Grenoble 38043, France}

\author{Yong-Hao Sun}
\affiliation{Institute of Physics, Chinese Academy of Sciences, 100190 Beijing, China}
\affiliation{School of Physical Sciences, University of Chinese Academy of Sciences, 100049 Beijing, China}

\author{Wei-Hua Wang}
\affiliation{Institute of Physics, Chinese Academy of Sciences, 100190 Beijing, China}
\affiliation{School of Physical Sciences, University of Chinese Academy of Sciences, 100049 Beijing, China}
\affiliation{Songshan Lake Materials Laboratory, Dongguan 523808, China}
\affiliation{Dongguan Institute of Materials Science and Technology, Chinese Academy of Sciences, Dongguan 523808, China}

\author{Ruta Beatrice}
\thanks{Contact author: \textnormal{beatrice.ruta@neel.cnrs.fr}}
\affiliation{Institut Néel, Université Grenoble Alpes and Centre National de la Recherche Scientifique, 25 rue des Martyrs - BP 166, 38042, Grenoble cedex 9, France}
\affiliation{European Synchrotron Radiation Facility, 71 avenue des Martyrs, CS 40220, Grenoble 38043, France}



\begin{abstract} 
Understanding the microscopic mechanisms underlying polyamorphic transitions requires simultaneous knowledge of thermodynamic and structural evolution, which has remained experimentally inaccessible. Here, we combine in situ chip-based fast differential scanning calorimetry with high-energy synchrotron X-ray diffraction to simultaneously probe the thermodynamic and structural evolution during the polyamorphic transition of a Pd–Ni–P metallic glass. Correlating heat-flow signals with the corresponding structural evolution during both the forward and reverse transformations reveals a direct relationship between thermodynamic evolution and medium-range structural ordering. Based on this relationship, we identify two distinct transition mechanisms. In the glassy state, the transition proceeds continuously through a progressive medium-range ordering, accompanied by enhanced thermal stability and strongly temperature-dependent kinetics. In contrast, in the equilibrium supercooled liquid, the transition proceeds through the coexistence of distinct amorphous states with only a weak temperature dependence of the kinetics. These findings show that atomic mobility governs the transformation pathway by coupling thermodynamic evolution to structural ordering, providing new insights into polyamorphic transitions in nonequilibrium amorphous systems.
\end{abstract}
                            
\maketitle

\section{Introduction}

Polyamorphism refers to transitions between distinct amorphous states of the same chemical composition, by analogy with phase transformations in crystalline materials. Liquid–liquid transitions (LLTs) are a typical manifestation of polyamorphism and have been widely reported in molecular~\cite{Tanaka2004,Debenedetti2020}, atomic~\cite{Henry2020,Wei2013}, and ionic liquids~\cite{Wojnarowska2022,Yao2023}. In the glassy state, the corresponding glass–glass transitions (GGTs) have been observed in amorphous ice~\cite{Johari1996}, silicon~\cite{Fan2024}, chalcogenide glasses~\cite{Fujita2023}, and metallic glasses (MGs)~\cite{Sheng2007}, which are more commonly induced by changes in pressure and involve transformations between low- and high-density amorphous states. Polyamorphism reveals a rich space for tuning disordered systems: at a fixed chemical composition, the same system can exhibit distinct amorphous states with different structures~\cite{Henry2020,Wei2013}, densities~\cite{Fan2024}, refractive indices~\cite{Tanaka2004}, enthalpies~\cite{Du2020}, and mechanical properties~\cite{Shen2020}. More fundamentally, polyamorphism reflects the intricate interplay between thermodynamics, atomic structure, and kinetics in disordered systems, providing important insights into the nature of amorphous matter and its transformations. Polyamorphism is therefore not only of potential practical relevance but also of central importance in glass and condensed-matter physics.

Metallic glasses are particularly attractive for studies of polyamorphism because (i) in the absence of directional covalent bonding, molecular chains, or ionic interactions, their atomic configurations possess a high degree of freedom~\cite{Wang2025,Gao2026}, and (ii) a rich dynamical pathways that thermally activated atomic rearrangements even in the glassy state~\cite{Zhao2025,Evenson2014,Shen2026}. To date, polyamorphic transitions have been reported in Pd–Ni–P~\cite{Lan2017,Du2020,Shen2022,Ouyang2023}, La(Ce)–Co–Al~\cite{Lou2020,Shen2020,Shen2023,Yang2026}, Zr–Cu–Al–Y~\cite{Jiang2025}, Zr–Ti–Cu–Ni–Be and Zr–Cu–Ni–Al–Nb~\cite{Stolpe2016}, and high-entropy MGs~\cite{Luan2022,Cao2024,Yang2024,Zhang2022}. These transitions typically manifest as an anomalous exothermic peak or an anomalous change in specific heat near the glass-transition temperature ($T_{\mathrm{g}}$) in calorimetric heating curves. Exceptions include Au–Cu–Si–Ag–Pd~\cite{Gallino2018,Wang2016,Hechler2018} and Pd–Ni–P~\cite{Shen2023Kinetics,daSilvaPinto2024} MGs, in which GGTs require prolonged annealing at sub-$T_{\mathrm{g}}$ temperatures, while some heating curves show no detectable LLT exotherm. Because such annealing-induced GGTs require active atomic rearrangement within a largely frozen glassy configuration, how polyamorphism proceeds at the microscopic level remains an intriguing yet open question.     

Fast differential scanning calorimetry (FDSC) provides access to thermodynamic signatures that are difficult to resolve under conventional heating conditions. By employing sufficiently high heating rates, the FDSC shifts the devitrification temperature far above the fictive temperature, $T_{\mathrm{f}}$, thus improving the glass-transition overshoot endotherm ~\cite{Wang2016,Monnier2020,Schawe2019,Zheng2019} and revealing hidden thermodynamic processes that are inaccessible under conventional heating conditions~\cite{Yang2020,Gao2020,Shen2025,Cangialosi2024}. In the case of polyamorphism, previous studies have shown that transformed states in both molecular systems and MGs exhibit pronounced glass-transition overshoots during rapid heating, indicating that these overshoots are likely associated with the reverse transition from the transformed amorphous state to the supercooled liquid (SCL)~\cite{Shen2022,Kobayashi2016}. Further tracking the evolution of the overshoot position and height with annealing time provides insights into the kinetics of polyamorphism and distinguishes between continuous and coexistence transition mechanisms~\cite{Cao2022,Shen2022,Kobayashi2016}. However, the structural origin of these changes in enthalpy on the atomic scale remains unknown. Moreover, structural aging and polyamorphic transition can occur simultaneously during annealing, making their respective thermodynamic contributions difficult to disentangle solely from calorimetric measurements. A simultaneous investigation of structural and thermodynamic evolution is therefore essential for clarifying their underlying mechanisms.

To address this challenge, we combine FDSC with high-energy synchrotron X-ray diffraction (XRD)~\cite{Romer2025,Stanko2024}, to directly correlate thermodynamic signatures with atomic-scale structural changes throughout the thermal protocol of polyamorphic transition in a Pd$_{41.5}$Ni$_{41.5}$P$_{17}$ MG (\textbf{Fig.~1a}). We resolve the structural evolution from the as-quenched state to the transformed amorphous state during annealing, and subsequently during the pronounced overshoot endotherm induced by rapid heating. By taking advantage of the complete overshoot signal and combining it with simultaneously obtained atomic structural information, we establish a quantitative relationship between structure and thermal stability, and identify two distinct transition mechanisms governing thermally induced polyamorphism. These findings provide a general framework that can be extended to other amorphous systems to investigate rapid nonequilibrium transitions, while underscoring the crucial role of atomic dynamics in driving such transitions.


\section{experimental methods}
\subsection{Sample preparation}

A Pd--Ni pre-alloy was first prepared by arc melting under an Ar atmosphere, followed by sealing it together with P in an Ar-filled quartz tube and induction melting to obtain the master alloy with a nominal composition of Pd$_{41.5}$Ni$_{41.5}$P$_{17}$ (at.\%). The master alloy was further purified in a B$_3$O$_2$ flux at 1273~K for 4~h. MG ribbons were produced by melt spinning under an argon atmosphere. A small sample with an estimated mass of 80~ng was cut from the ribbon and pre-melted on the FDSC chip for subsequent FDSC and XRD experiments.

\subsection{Thermodynamic measurements}

Thermodynamic measurements were performed using a Mettler-Toledo Flash DSC 2+ calorimeter. FDSC measurements were conducted under a flowing N$_2$ atmosphere at 50~mL~min$^{-1}$. The sample ($\sim$20~$\mu$m in diameter) was preheated to 1023~K (which is above the liquidus temperature for melting) and subsequently cooled at 10,000~K~s$^{-1}$ to different temperatures for annealing. All FDSC data presented in this study were obtained during subsequent heating scans following multiple independent repetitions of the annealing protocol, with varying annealing time, $t_{\mathrm{a}}$, and annealing temperature, $T_{\mathrm{a}}$. Heating rates ranging from 50 to 1000~K~s$^{-1}$ were employed. Conventional DSC measurements were carried out using a Mettler-Toledo DSC3 calorimeter. The samples included as-cast ribbons and samples annealed at 563~K for 36,000~s, with each specimen having a mass of $\sim$10~mg. All measurements were conducted at a heating rate of 10~K~min$^{-1}$.

\subsection{High-energy XRD measurements}

High-energy XRD measurements were carried out at the ID15A beamline of the European Synchrotron Radiation Facility (ESRF) in Grenoble, France. X-ray scattering patterns were collected using a hybrid photon-counting pixel detector (Pilatus3 X CdTe), with a fastest acquisition time of 0.1~s per frame. During heating measurements, a heating rate of 50~K~s$^{-1}$ was applied to ensure sufficient data acquisition, corresponding to a temperature increment of 5~K per frame. During annealing measurements, data were acquired with an acquisition time of 0.1--1~s per frame. The X-ray photon energy was fixed at 67.808~keV, and the X-ray beam was focused to $1\times2~\mu$m$^2$ (FWHM, H$\times$V), where H and V denote the horizontal and vertical directions, respectively.

Azimuthal integration of the diffraction patterns was performed using the pyFAI library, which accounts for the X-ray polarization factor and the solid angle subtended by each detector pixel. The experimental geometry was calibrated using CeO$_2$ powder spread onto the chip surface, and the sample-to-detector distance was determined to be 199.3~mm. The valid $q$-range was 0.45--18~\AA$^{-1}$, with the lower limit defined by the 2~mm-diameter beamstop placed near the exit Kapton window. The $S(q)$ was calculated from the measured diffraction data using the PDFgetX2 program~\cite{Qiu2004}.

\subsection{In Situ FDSC-XRD Setup}

A custom-designed FDSC chamber was mounted on the ESRF ID15A beamline, as shown in \textbf{Fig.~1a}. The chamber was provided by the Disordered Systems research group at the University of Padua, and a detailed description of the setup can be found in Ref.~\cite{Sun2025}. The chamber was maintained under high-vacuum conditions ($\sim10^{-5}$~bar) to minimize X-ray scattering from air. To further minimize the crystalline signal from the Au alloy of the heating pan and coil in the FDSC sensor, the MG sample was pre-melted in the gap between them, thereby preventing excessive Au signal from being collected. The additional background mainly originated from the amorphous SiN membrane and was negligible~\cite{Sun2025}. As verified by X-ray line scans, the sample displacement in the vertical direction remained within 1~$\mu$m during repeated melt-quench cycles.

To synchronize the X-ray shutter, detector, and FDSC temperature protocol, all instructions were integrated into the ESRF BLISS system and could be executed using a single macro. Considering the $\sim$3--5~s delay between issuing and executing the FDSC temperature program, an additional 10~s waiting period at room temperature was introduced between issuing the temperature command and opening the X-ray shutter. The shutter opening time, marked by abrupt changes in the chip-recorded sample temperature and heat-flow signals (\textbf{Fig.~1a}), was defined as the starting time, $t_0$, of the temperature program. This approach ensured synchronization between the sample temperature and diffraction data acquisition (\textbf{Fig.~1b}).

To minimize local heating induced by X-ray irradiation, we used an attenuator to reduce the beam transmission to 10\%, limiting the additional temperature increase to approximately 3~K. Accordingly, the temperatures in the FDSC annealing protocol were corrected by $-3$~K, enabling direct comparison between the thermodynamic and structural data. This temperature correction was estimated by comparing the heat-flow curves with and without X-ray irradiation, consistent with estimates based on diffraction data~\cite{Sun2025}.

\section{Results}
\subsection{Isothermal XRD and FDSC Measurements}

\textbf{Figure~1b} presents the structural evolution during isothermal annealing at 563~K after melt-quenching at a rate of 10,000~K~s$^{-1}$. Compared with the state at an annealing time $t_{\mathrm{a}} = 1$~s, the first sharp diffraction peak (FSDP) of the structure factor, $S(q)$, where $q$ is the scattering vector, remains essentially unchanged after $t_{\mathrm{a}} = 100$~s. In contrast, at $t_{\mathrm{a}} = 5000$~s and 30,000~s, the FSDP becomes noticeably narrower and more intense (a magnified view of the FSDP profiles is provided in \textbf{Fig.~S1}), revealing enhanced medium-range order associated with the polyamorphic glass--glass transition~\cite{Lan2017,Du2020,Shen2022,Ouyang2023}. For simplicity, the transformed glass is hereafter denoted as $g_{2}$, whereas the as-quenched glass is denoted as $g_{1}$.

The corresponding FDSC curves (\textbf{Fig.~1c}), obtained at different $t_{\mathrm{a}}$ (see Experimental Methods Section), exhibit a closely correlated thermodynamic evolution. At $t_{\mathrm{a}} = 100$~s, a single endothermic peak characteristic of structural aging is observed. At $t_{\mathrm{a}} = 5,000$~s, a second endothermic peak emerges and partially overlaps with the aging peak, signaling the polyamorphic transition. Once the transition is complete (e.g., $t_{\mathrm{a}} = 30,000$~s), the low-temperature peak disappears at the expense of the high-temperature peak, and $T_{\mathrm{g}}$---defined as the intersection of the glass baseline and the tangent to the rising heat-flow step---increases by 42~K compared with the as-quenched $g_{1}$, consistent with previous reports~\cite{Du2020,Shen2022}. Notably, this overshoot is resolved only under rapid heating and is therefore a consequence of the high heating rate. At a conventional heating rate of 10~K~min$^{-1}$, such a clear overshoot is not observed, despite the disappearance of the exothermic peak in the SCL region after annealing at 563~K for 36,000~s (\textbf{Fig.~S2}). 

\begin{figure*}[t]
    \centering   
     \includegraphics[width=0.7\textwidth]{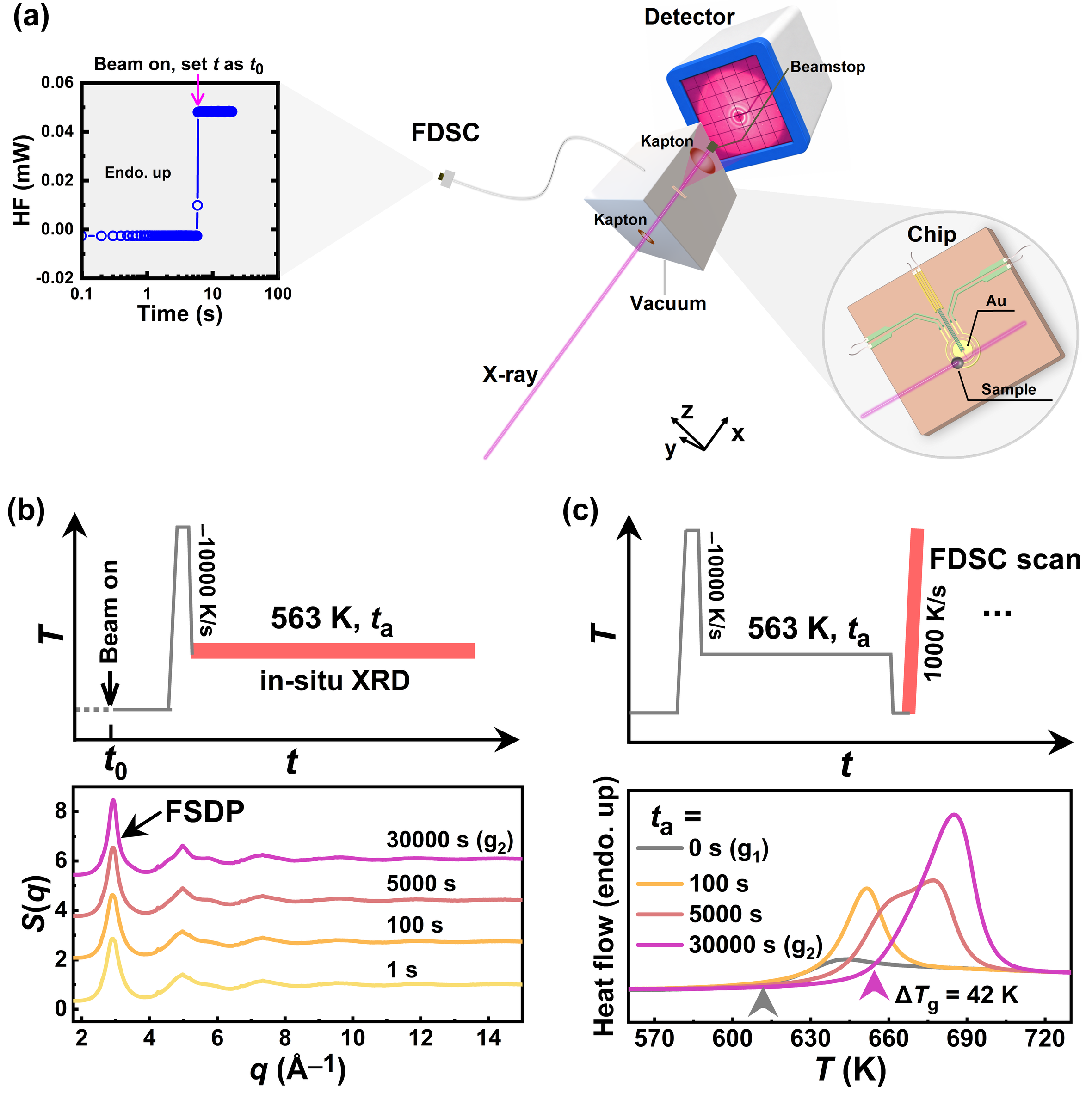}
    \caption{Integrated FDSC--XRD measurement and experimental protocols. 
    (a) Simplified sketch of the integrated FDSC--XRD setup. The system consists of a high-temperature FDSC chip, a high-energy X-ray beam, an MG sample, and an X-ray detector. To avoid diffraction signals from the heating resistor (i.e., the plate and coil), the sample was melt-quenched in the gap between them. Upon opening the X-ray shutter, irradiation induces an endothermic signal, the onset of which defines the starting point $t_0$ of the XRD temperature protocol. 
    (b) Structural evolution during isothermal annealing at 563~K after melt-quenching. The upper panel shows the temperature--time protocol, and the lower panel shows the time-resolved evolution of $S(q)$ at different annealing times $t_{\mathrm{a}}$. 
    (c) Thermodynamic evolution during annealing at 563~K. The upper panel shows the temperature--time protocol, including melting, quenching, annealing for different $t_{\mathrm{a}}$, and subsequent heating at 1000~K~s$^{-1}$. The lower panel shows the corresponding heat-flow curves, with each curve representing an independent experiment at a different $t_{\mathrm{a}}$. For the as-quenched glass ($g_{1}$), $T_{\mathrm{g}}$ ($t_{\mathrm{a}}=0$~s) is 612~K (grey arrow), whereas for the transformed glass ($g_{2}$), $T_{\mathrm{g}}$ ($t_{\mathrm{a}}=30,000$~s) is 654~K (pink arrow).}
\end{figure*}

To establish a quantitative relationship between thermodynamic and atomic structure evolution, \textbf{Fig.~2} presents the temporal evolution with $t_{\mathrm{a}}$ of the fictive temperature,~$T_{\mathrm{f}}$, the cube of the reciprocal of the FSDP center position, $(1/q_1)^3$, and the full width at half maximum (FWHM) of the FSDP at 563~K. Here, $T_{\mathrm{f}}$, which characterizes the thermodynamic state of the sample, was determined using the area-matching method~\cite{Moynihan1976}; the parameter $(1/q_1)^3$ is related to the atomic packing configuration~\cite{Yavari2005}, while the FWHM reflects the degree of medium-range structural ordering~\cite{Ma2009}.

The evolution proceeds through two distinct regimes. During the initial stage ($t_{\mathrm{a}}\lesssim600$~s), the sample undergoes aging, as evidenced by a single endothermic peak in the FDSC curves and a decrease in $T_{\mathrm{f}}$ by approximately 47~K, approaching equilibrium (i.e., $T_{\mathrm{f}}$ shifts toward the annealing temperature of 563 K)~\cite{Chen2023,Xing2026,Ruta2017,daSilvaPinto2024}. Meanwhile, both $(1/q_1)^3$ and FWHM decrease only slightly, indicating limited structural rearrangements during aging. To quantitatively characterize the aging kinetics, we fitted the experimental data using the Kohlrausch--Williams--Watts (KWW) function,

\begin{equation}
y(t_{\mathrm{a}})
=
A
+
D
\exp\left[
-\left(
\frac{t_{\mathrm{a}}}{\tau}
\right)^{\beta}
\right],
\end{equation}

where $A$ and $D$ are constants describing the initial and final values of the relaxing quantity, respectively, $\tau$ is the characteristic relaxation time, and $\beta$ is the shape exponent. A global KWW fit was simultaneously performed for $(1/q_1)^3$ and FWHM. The global fit yields a shared stretching exponent of
$\beta = 0.53 \pm 0.02$ and a shared characteristic relaxation time of $\tau = 323.8 \pm 8.7$~s. 

Beyond 600~s, the emergence of a second endothermic peak in the FDSC curve marks the onset of the polyamorphic transition. This second stage is characterized by a much more pronounced reduction in both $(1/q_1)^3$ and FWHM, reflecting enhanced medium-range ordering and atomic rearrangements, together with an additional decrease of $\sim$50~K in $T_{\mathrm{f}}$ over a narrow time interval, reaching a value much lower than the applied $T_{\mathrm{a}}$. These correlated thermodynamic and structural changes indicate the continuous evolution from $g_{1}$ to $g_{2}$. When $t_{\mathrm{a}}$ approaches $\sim$7000~s, the rate of change of all parameters slows markedly, indicating that the polyamorphic transition is almost complete and that the subsequent slow evolution of the system is likely dominated by aging of the newly formed $g_{2}$, although residual contributions from the transition itself cannot be ruled out~\cite{Shimizu2014}.

\begin{figure}[t]
    \centering   
     \includegraphics[width=0.48\textwidth]{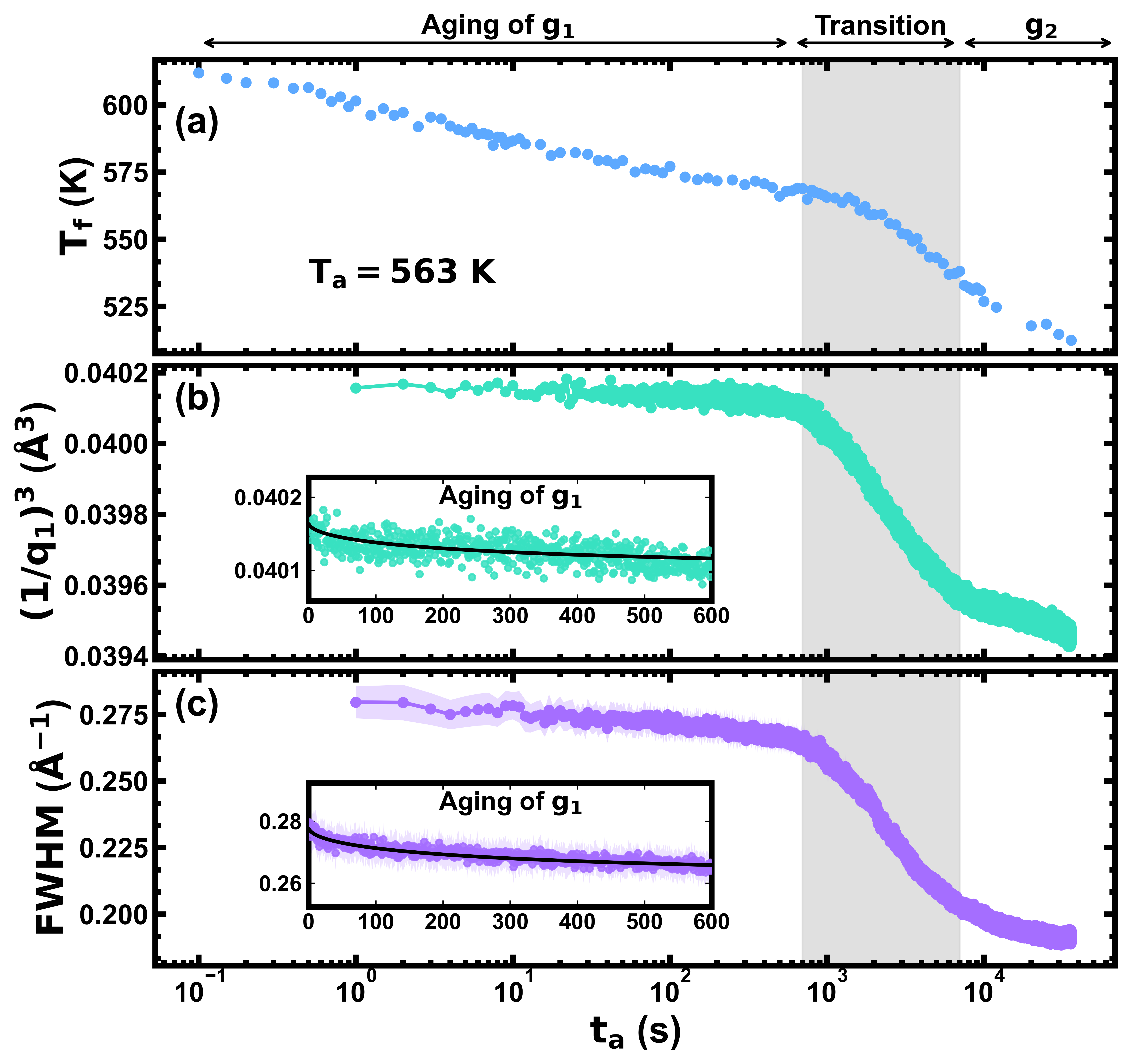}
    \caption{Correlation between thermodynamic and structural evolution during annealing at $T_{\mathrm{a}}=563$~K. The top, middle, and bottom panels show the evolution of $T_{\mathrm{f}}$~(a), $(1/q_1)^3$~(b), and the FWHM of the FSDP~(c), respectively. The gray-shaded region denotes the polyamorphic transition regime, during which $T_{\mathrm{f}}$ decreases by $\sim$50~K~(a), accompanied by increased structural ordering~(b,c). By contrast, during the initial aging regime, $T_{\mathrm{f}}$ decreases more smoothly and approaches the equilibrium plateau corresponding to $T_{\mathrm{a}}$, whereas the structural parameters exhibit only minor changes (see insets, where the black solid lines represent the global KWW fits to the $(1/q_1)^3$ and FWHM data; X-axis: $t_{\mathrm{a}}$, Y-axis: $(1/q_1)^3$ or FWHM). The shaded purple region in~(c) corresponds to the fitting uncertainties obtained from the bi-Gaussian fitting of the FSDP.}
\end{figure}

\subsection{Fast-Heating XRD and Calorimetry Measurements}

To investigate structural evolution during rapid heating, $g_{1}$ and $g_{2}$ ($T_{\mathrm{a}}=583$~K and $t_{\mathrm{a}}=3,600$~s, and subsequently cooled to ambient temperature) were heated at a rate of 50~K~s$^{-1}$ while simultaneously recording their structural evolution in situ using XRD and their thermodynamic profile with FDSC (see Methods).

During heating, $g_{1}$ sequentially undergoes glass transition, crystallization, and crystal melting, all of which are clearly reflected in the heat-flow signal (see \textbf{Fig.~3a}). The XRD patterns distinctly capture the crystallization and melting processes, manifested by the appearance and subsequent disappearance of Bragg peaks, respectively.

\textbf{Figure~3b} shows that, during heating, the transformed glass $g_{2}$ also undergoes crystallization followed by melting, as consistently reflected by both the heat-flow signal and the XRD patterns. In particular, a pronounced post-$T_{\mathrm{g}}$ endotherm is observed in the heat-flow curve after the glass transition, overlapping in temperature with the pronounced broadening of the FSDP. This structural change is the exact opposite of the FSDP narrowing previously observed during the annealing-induced transition from $g_{1}$ to $g_{2}$ (\textbf{Fig.~2c}), demonstrating that the endotherm corresponds to the reversal of the structural ordering established during annealing. Furthermore, this reverse structural evolution cannot be suppressed by rapid heating, as further confirmed by additional calorimetric measurements (\textbf{Fig.~S3}).

\begin{figure*}[!t]
    \centering
    \includegraphics[width=0.65\textwidth]{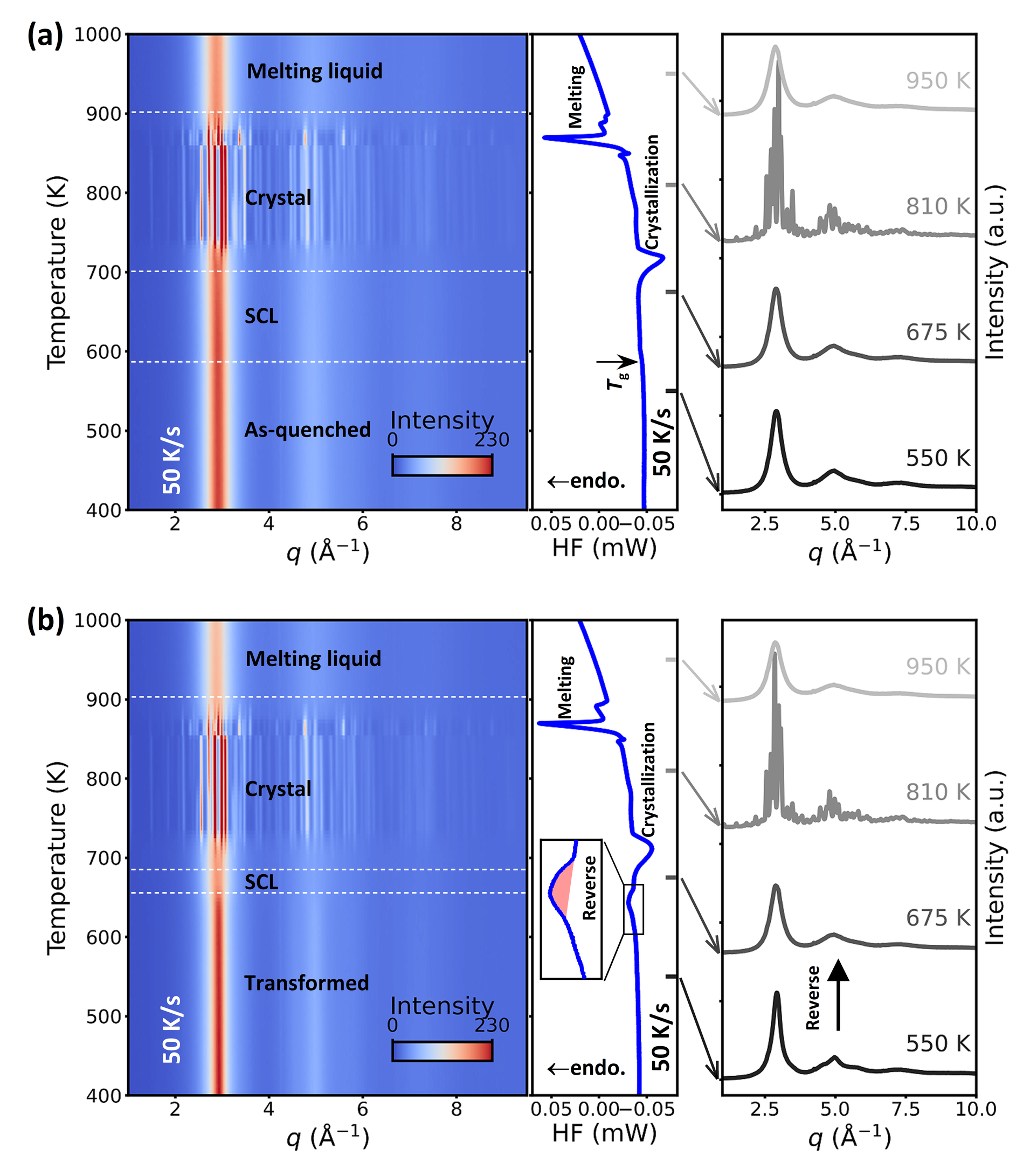}
    \caption{Evolution of the as-quenched sample and the transformed sample $g_{2}$ ($T_{\mathrm{a}}=583$~K, $t_{\mathrm{a}}=3600$~s) during heating at a rate of 50~K~s$^{-1}$. (a) The as-quenched sample (quenched at 10,000~K~s$^{-1}$) undergoes a glass transition, crystallization, and subsequent crystal melting upon heating. (b) The transformed sample $g_{2}$ undergoes a reverse transition to the SCL, followed by crystallization and subsequent crystal melting. The left panels show the temperature-dependent XRD intensity, with the same temperature axis as the heat-flow curves in the middle panels. The inset endothermic peak in (b) corresponds to the reverse transition from $g_{2}$ to the SCL. The right panels show XRD patterns collected at representative temperatures corresponding to the transformed state, SCL, crystalline state, and molten liquid, respectively. The XRD and heat-flow data were acquired simultaneously.}
    \label{fig:Figure3}
\end{figure*}

\textbf{Figure~4} summarizes the temperature dependence of the FSDP position, $q_1$, and its FWHM during heating of $g_{1}$ and $g_{2}$. The temperature range of 680--905~K (highlighted in purple in~\textbf{Fig.~4a~and~4b}) corresponds to the crystalline state and was therefore excluded from the analysis.

The decrease in $q_1$ with increasing temperature reflects the thermal expansion of the MGs~\cite{Yavari2005}. For $g_{1}$, $q_1$ decreases monotonically with increasing temperature, exhibiting typical thermal expansion at low temperatures, with $\mathrm{d}q_1/\mathrm{d}T \approx -4.02\times10^{-5}$~\AA$^{-1}$~K$^{-1}$, consistent with the value reported for compositionally similar MGs~\cite{Yavari2005}. Above $\sim$540~K, the temperature dependence becomes less pronounced, consistent with the progressive annihilation of excess free volume in the rapidly quenched glass~\cite{Yavari2005,Duan2024}. After the glass transition, the sample enters the SCL state, where the thermal expansion becomes more pronounced, with $\mathrm{d}q_1/\mathrm{d}T \approx -8.86\times10^{-5}$~\AA$^{-1}$~K$^{-1}$, reflecting the additional contribution from configurational motions in the liquid  ~\cite{Lunkenheimer2023,Lu2023}. With further increasing temperature, the system undergoes crystallization and subsequent melting. Notably, the $q_1$ values of the equilibrium melt fall on the linear extrapolation of the SCL data, indicating structural continuity between the SCL and the high-temperature equilibrium liquid.

In contrast, $g_{2}$ exhibits systematically larger $q_1$ values than $g_{1}$, indicating a different atomic packing. Nevertheless, both states exhibit nearly identical thermal expansion in the glassy regime, with comparable values of $\mathrm{d}q_1/\mathrm{d}T$. Unlike $g_{1}$, $g_{2}$ shows no evident signature of free-volume annihilation during heating, and $q_1$ follows the same temperature dependence throughout the entire glassy range. At $\sim$640~K, $q_1$ undergoes an abrupt decrease before merging with the SCL branch followed by $g_{1}$, demonstrating that the transformed glass $g_{2}$ reverses back to the same liquid state as $g_{1}$ prior to crystallization.

\textbf{Figure~4b} shows the temperature dependence of the FWHM of the FSDP. In contrast to the evolution of $q_1$, the FWHM remains nearly constant throughout the glassy state and increases approximately linearly upon entering the SCL, consistent with previous scattering studies attributing this behavior to the freezing of medium-range order below $T_{\mathrm{g}}$~\cite{Ryu2025}. Strikingly, $g_{2}$ exhibits a pronounced increase in the FWHM in the transition region, ultimately converging, like $q_1$, with the SCL branch of $g_{1}$. This observation provides direct structural evidence that the reverse transformation restores the medium-range structural state of the original SCL.

To quantify the thermodynamic reversibility of the transformation, a dedicated temperature protocol was designed. The $g_{2}$ sample, obtained by annealing at 563~K for 36,000~s, was heated at 50~K~s$^{-1}$ to different quenching temperatures, $T_{\mathrm{q}}$, in the range of 598--678~K, rapidly quenched to room temperature, and subsequently reheated. The resulting heat-flow curves therefore reflect the thermodynamic state frozen at each $T_{\mathrm{q}}$.

As shown in \textbf{Fig.~4c}, increasing $T_{\mathrm{q}}$ drives a continuous evolution of the calorimetric response from the characteristic $g_{2}$ behavior toward that of $g_{1}$, manifested by the progressive disappearance of the overshoot endotherm and the recovery of the low-temperature glass-transition step. This evolution is quantified in \textbf{Fig.~4d}. $T_{\mathrm{f}}$ remains nearly constant up to $T_{\mathrm{q}}\sim620$~K, before increasing continuously as thermal activation drives the glass toward the equilibrium SCL state. For $T_{\mathrm{q}}\geq663$~K, $T_{\mathrm{f}}$ becomes identical to that of a glass quenched from the high-temperature melt, demonstrating that the system has fully recovered the ordinary $g_{1}$ state. As expected, the $g_2$-to-SCL reverse transition exhibits a pronounced heating-rate dependence, with its temperature range progressively shifting toward higher temperatures as the heating rate increases. However, the overall enthalpy change remains unchanged, even at the fastest heating rate of 5,000~K~s$^{-1}$ (\textbf{Fig.~S3}). Combined with the isothermal annealing experiment (\textbf{Fig.~2}), these results establish the reversibility of the $g_{1}$--$g_{2}$ transition and provide a direct quantitative correlation between thermodynamic and structural evolution.

\begin{figure*}[t]
    \centering   
      \includegraphics[width=0.7\textwidth]{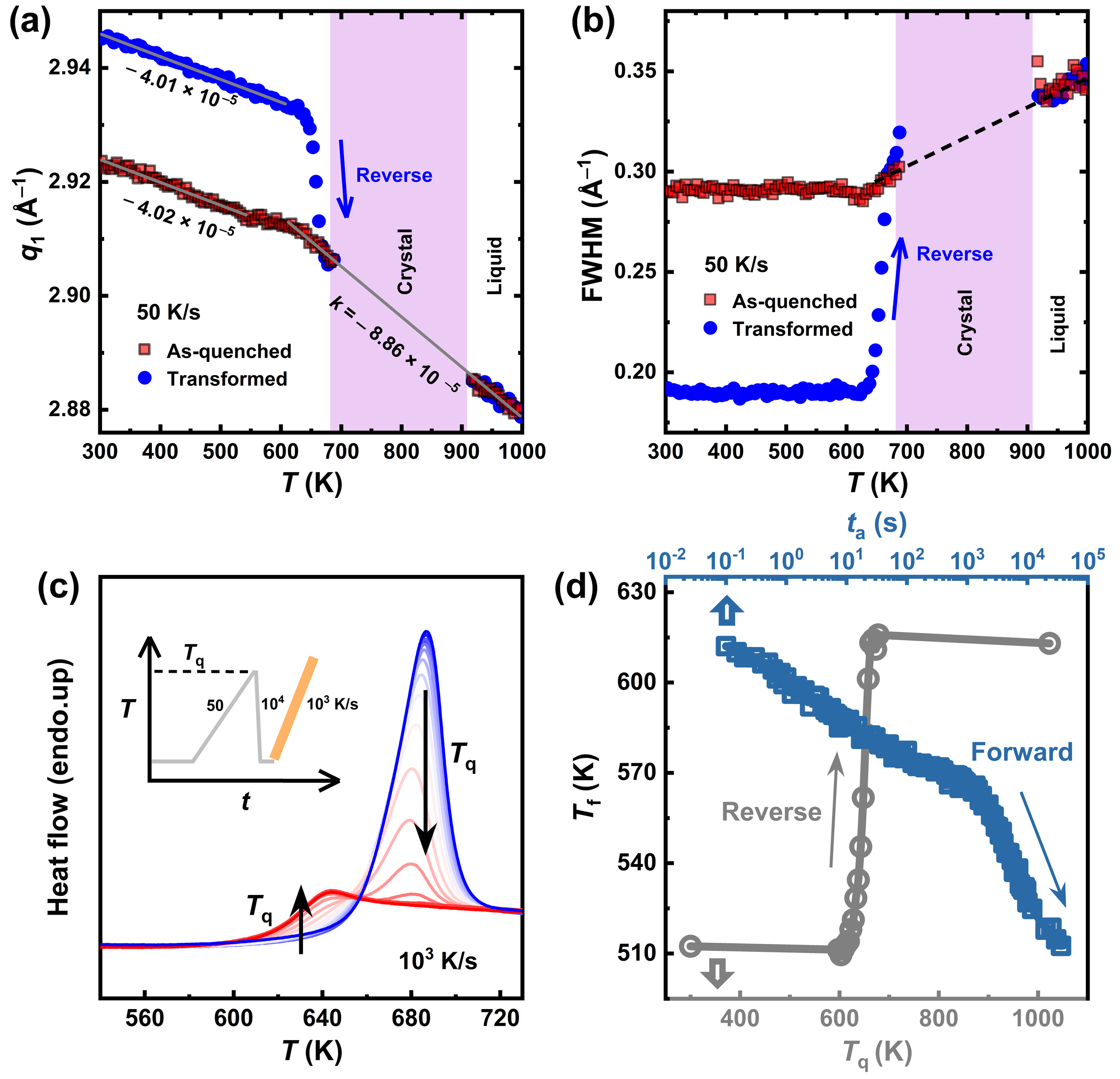} 
    \caption{Reversible structural and thermodynamic transition during rapid heating. Temperature dependence of $q_1$~(a) and the FWHM of the FSDP~(b) for the as-quenched ($g_{1}$) and transformed ($g_{2}$) glasses during heating at 50~K~s$^{-1}$. The reported $q_1$ and FWHM values represent averages over four independent measurements. Solid lines in~(a) are linear fits to the $q_1$ data in the glassy and SCL regimes. (c) Heat-flow curves of $g_{2}$, obtained by annealing at 563~K for 36,000~s, after interruption of the reverse transition at different quenching temperatures, $T_{\mathrm{q}}$ (598--678~K). The curves are colored from blue to red with increasing $T_{\mathrm{q}}$. The inset shows the temperature protocol: $g_{2}$ was first heated at 50~K~s$^{-1}$ to different $T_{\mathrm{q}}$, rapidly quenched at 10,000~K~s$^{-1}$, and subsequently reheated at 1000~K~s$^{-1}$ while recording the heat flow. (d) $T_{\mathrm{f}}$ as a function of $T_{\mathrm{q}}$, representing the reverse process. For comparison, $T_{\mathrm{f}}$ as a function of $t_{\mathrm{a}}$ during annealing at 563~K is also included, representing the forward transition shown in Fig.~2a. A zoomed-in view of the $T_{\mathrm{q}}$ axis around the reverse transition is provided in \textbf{Fig.~S4}.}
\end{figure*}

\subsection{Continuous and Coexisting Mechanisms}
We next examine the influence of annealing temperature on the transformation behavior by increasing $T_a$ from 563 to 623~K. In contrast to the previously studied case of $T_a = 563$~K, at 623~K the annealing temperature exceeds $T_f$ of the as-quenched state, allowing the polyamorphic transition to occur under liquid-state conditions, where the $\alpha$-relaxation is fully activated~\cite{Monnier2020,Xing2024}.

As shown in \textbf{Fig.~5a}, when $T_{\mathrm{a}}=563$~K, the system undergoes normal aging prior to 600~s, during which only the first endothermic peak is observed in the FDSC curve. Once the polyamorphic transition begins ($t_{\mathrm{a}}>600$~s), a second endothermic peak gradually emerges, while the first peak continuously shifts toward higher temperatures before eventually merging with the second. Simultaneously, the first endothermic peak progressively broadens and its intensity decreases. This continuous evolution demonstrates that, during the GGT process, the aged $g_{1}$ state does not transform directly into $g_{2}$, but instead evolves through a series of intermediate glassy states connecting $g_{1}$ and $g_{2}$. Meanwhile, the second endothermic peak increases in magnitude and shifts continuously to higher temperatures, indicating that the reverse transition of these intermediate states into the SCL follows the same continuous pathway. Consistent with the FDSC results, the XRD data reveal a progressive narrowing of the FSDP that coincides with the emergence of the second endothermic peak, confirming that the structural evolution proceeds entirely within the amorphous state without detectable crystallization.

In contrast, upon annealing at $T_{\mathrm{a}}=623$~K (\textbf{Fig.~5b}), the sample remains in the equilibrium SCL state, with no observable enthalpy aging. However, after $\sim$3~s, the polyamorphic transition begins, marked by the emergence of a second endothermic peak in the FDSC curve. Unlike the behavior at 563~K, the position of the first endothermic peak remains unchanged throughout the transition, while its height gradually decreases. Simultaneously, the second endothermic peak grows in magnitude without shifting in temperature. These observations demonstrate that the transition proceeds primarily through the coexistence of two distinct amorphous states rather than through a continuum of intermediate states~\cite{Cao2024,Cao2022,Shen2022}. In this regime, the SCL partially transforms directly into the thermodynamically more stable $g_{2}$ glass. After approximately 40~s, crystallization sets in, as evidenced by the appearance of Bragg reflections in the XRD patterns and the concurrent reduction in the area of the second endothermic peak in the FDSC curve.

The coexistence of two amorphous states may originate from compositional phase separation~\cite{Kim2013}. However, this scenario can be ruled out based on two independent observations. First, the $T_{\mathrm{g}}$ of $g_{1}$ remains unchanged throughout the transition, whereas compositional variation would be expected to shift the $T_{\mathrm{g}}$ values of the two separated glasses. Second, no sub-rings are observed in the two-dimensional diffraction patterns, in contrast to the signatures reported for phase-separated MGs~\cite{Jiang2025,Kim2013}.

\begin{figure*}[t]
    \centering   
    \includegraphics[width=0.9\textwidth]{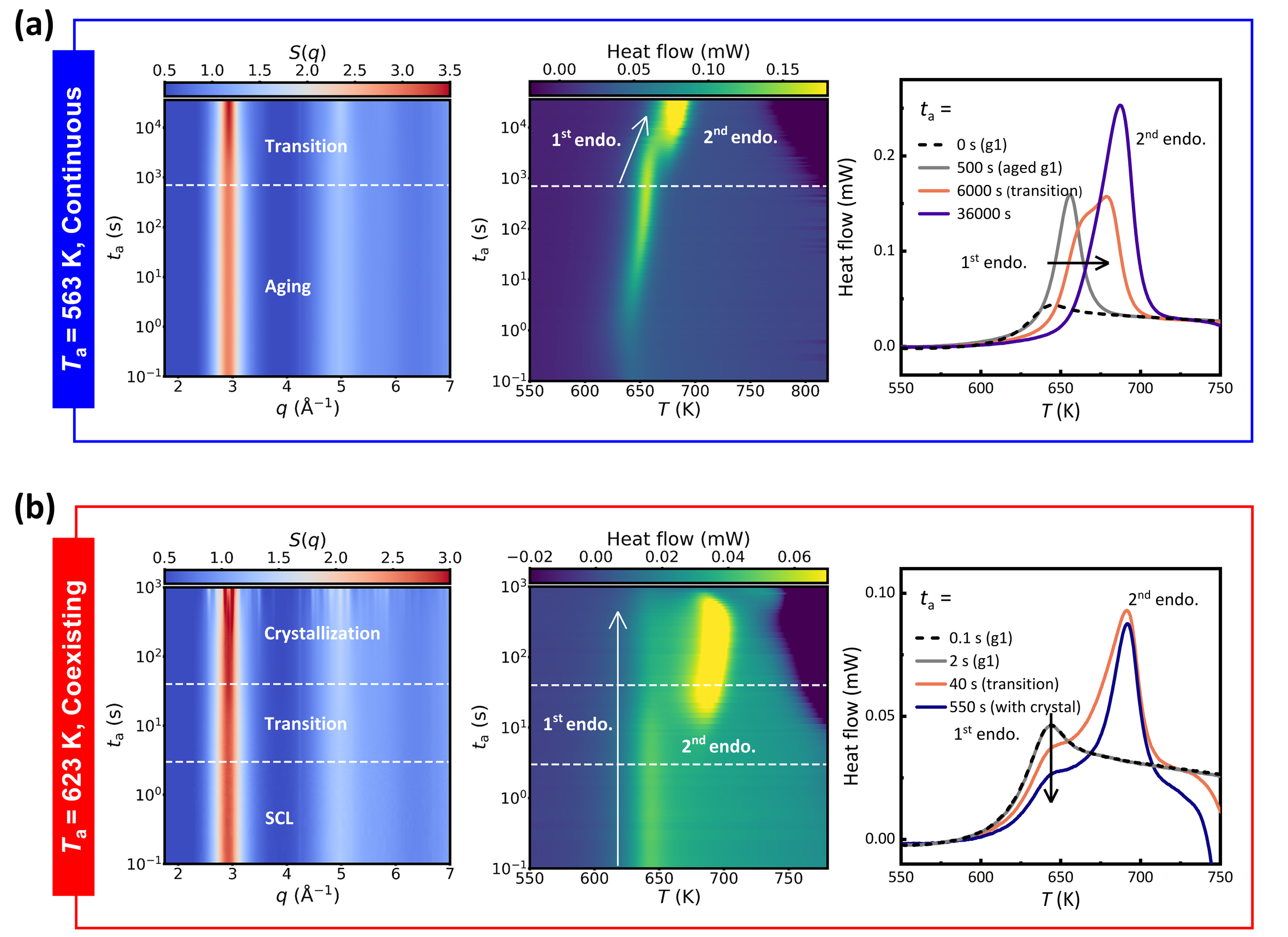} 
    \caption{Continuous and coexisting mechanisms. (a) XRD and FDSC results at $T_{\mathrm{a}}=563$~K, where the sample undergoes aging followed by the GGT process. The FDSC curves, measured at 1000~K~s$^{-1}$, exhibit two partially overlapping endothermic peaks during the GGT, with the first peak progressively shifting toward higher temperatures. (b) XRD and FDSC results at $T_{\mathrm{a}}=623$~K, where the sample sequentially passes through the thermodynamic equilibrium SCL state, the transition to $g_{2}$, and crystallization. The FDSC curves consistently show two well-separated endothermic peaks during the transition; the position of the first peak remains unchanged, while its intensity gradually decreases.}
\end{figure*}

The reverse-transition experiments demonstrate that the SCL, corresponding to the liquid state of $g_{1}$, is the thermodynamically stable high-temperature phase (\textbf{Fig.~4}). Consistent with this thermodynamic picture, 623~K should lie within the coexistence region of $g_{2}$ and the SCL rather than the stability region of either single phase. This implies that, even after prolonged annealing, a finite fraction of the SCL persists, preventing the complete transition to $g_{2}$. In contrast, 563~K lies within the thermodynamic stability region of $g_{2}$, enabling the system to transform completely into $g_{2}$ after sufficiently long annealing (\textbf{Fig.~5}).

From a structural perspective, the diffraction pattern in the coexistence regime arises from the superposition of the individual FSDP of $g_{2}$ and the SCL. However, because their peak positions nearly overlap ($q_1$ differs by less than 0.03~\AA$^{-1}$), the two contributions cannot be resolved independently in experiments and instead appear as a single FSDP.

\section{DISCUSSION}

The above results reveal a crossover temperature range separating two distinct microscopic mechanisms governing thermally induced polyamorphism. Below this crossover range, represented by 563~K, the transition proceeds continuously through a continuum of intermediate amorphous states. Above this range, as observed at 623~K, the transition occurs through the coexistence of two distinct amorphous states (\textbf{Fig.~5}). The transition mechanism can be readily distinguished from the evolution of the first endothermic peak in the FDSC curves: a progressive shift of the peak with annealing time signifies a continuous transition, whereas an invariant peak position accompanied by a gradual loss of intensity indicates a coexistence mechanism~\cite{Cao2022,Shen2022,Kobayashi2016}. 

\textbf{Figure~6} summarizes the temporal evolution of $T_{\mathrm{f}}$ at different annealing temperatures, together with the time of the polyamorphic transition, $t_{\mathrm{polyam.}}$. The temperature dependence of the FWHM (\textbf{Fig.~S5}) closely follows that of $T_{\mathrm{f}}$, demonstrating that the thermodynamic and structural signatures are governed by the same underlying kinetic process. Thus, the $t_{\mathrm{polyam.}}$ at each $T_{\mathrm{a}}$ is defined as the time at which the FWHM reaches 50\% of its total change. The data suggest two distinct kinetic regimes across the vitrification of the SCL (\(T_{\mathrm{f}}=613\) K, corresponding to a cooling rate of \(10{,}000\ \mathrm{K\,s^{-1}}\)). Below this temperature, the transition exhibits a strong temperature dependence, whereas above it, the kinetics are much less temperature-dependent, suggesting a substantially lower activation energy. Fitting all data below 608 K yields an activation energy of approximately \(281\ \mathrm{kJ\,mol^{-1}}\).

Previous studies on the Au-based MG-forming liquid have shown that a fragile-to-strong LLT induced by cooling from the melt leads to hierarchical aging in the resulting glassy states, characterized by progressively increasing activation energies, with the final aging stage exhibiting an activation energy consistent with that of the $\alpha$-relaxation in the strong liquid~\cite{Gallino2018,Monnier2020}. In our system, however, enthalpy relaxation (aging) and thermally induced polyamorphism are found to be two independent processes (\textbf{Fig.~2}). The equilibrium aging time, $t_{\mathrm{eq}}$, defined as the time at which the area of the first endothermic peak in the FDSC curve reaches saturation, exhibits a single Arrhenius-like temperature dependence across the entire temperature range. The corresponding activation energy, approximately 485~kJ~mol$^{-1}$, is in excellent agreement with the measurements based on enthalpy-relaxation experiments~\cite{Duan2024}. In contrast, the polyamorphism exhibits two distinct temperature-dependent regimes, corresponding to the coexisting and continuous transitions, with both activation energies visibly lower than that of aging (\textbf{Fig.~6}).

\begin{figure*}[t]
    \centering   
     \includegraphics[width=0.9\textwidth]{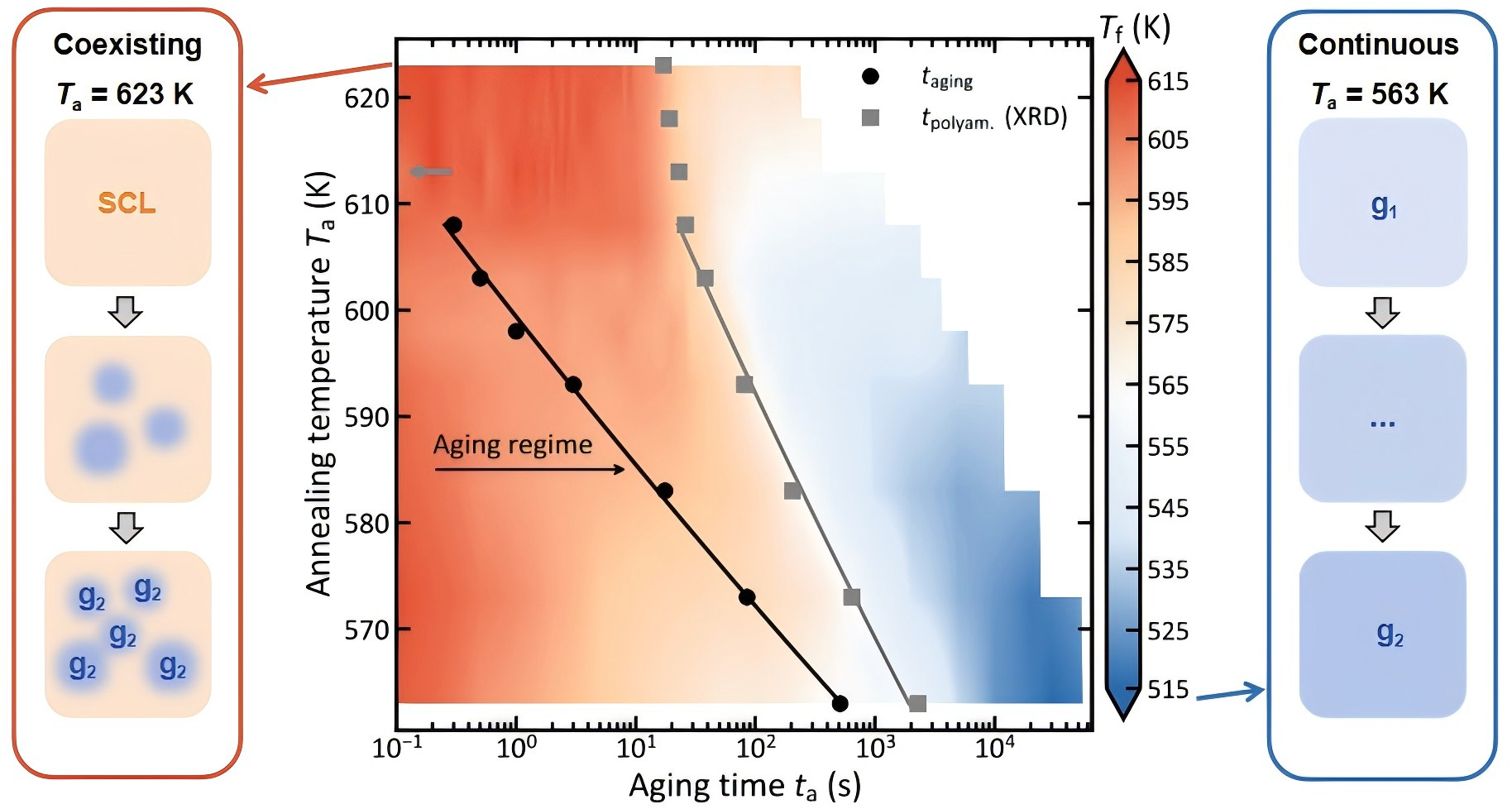} 
    \caption{Temperature dependence and mechanisms of the polyamorphic transition. Evolution of $T_{\mathrm{f}}$ with annealing time over the investigated temperature range. Black circles denote the equilibrium time for enthalpy relaxation, $t_{\mathrm{aging}}$. Gray squares indicate the characteristic time of the polyamorphic transition, $t_{\mathrm{polyam}}$, determined as the time at which the FWHM of the FSDP reaches 50\% of its total change (see \textbf{Fig.~S5}). The gray arrow indicates the fictive temperature ($T_{\mathrm{f}} = 613$~K) corresponding to a cooling rate of $10{,}000$~K~s$^{-1}$. Solid lines represent the corresponding Arrhenius fits. The schematic illustrates the two mechanisms: (i) nucleation and growth of discrete transformed regions ($g_{2}$, indicated by blue spots in the left column), which eventually coexist with the SCL; and (ii) continuous transformation of the glass, with increasing color intensity representing the progressive transformation from $g_{1}$ to $g_{2}$ (right column).}
\end{figure*}

A recent study by Rodriguez-Viejo \textit{et al.} has shown that the transition of an ultrastable molecular glass, or a deeply annealed glass, into the SCL is governed by the competition between the characteristic relaxation time of the glass, $\tau_{\mathrm{glass}}$, and the liquid $\alpha$-relaxation time, $\tau_{\alpha}$~\cite{VilaCosta2023}. When $\tau_{\mathrm{glass}}/\tau_{\alpha}$ is large, the transition proceeds through the coexistence of the two amorphous states, whereas comparable timescales lead to a cooperative, continuous transition. These findings suggest that the continuous and coexistence mechanisms are not material specific, but rather reflect a general behavior governed by kinetic constraints in glasses.

Our results support this general picture and further show that the polyamorphic transition is closely related to atomic mobility. While Rodriguez-Viejo \textit{et al.} revealed the transition from the glassy state to the SCL~\cite{VilaCosta2023}, this work reveals the transition from the equilibrium SCL or $g_{1}$ to $g_{2}$. Although these transitions involve different initial states, they share a similar underlying mechanism: when one system is in a SCL state, the absence of strong kinetic constraints renders the relative stability of local structural configurations highly sensitive to small energy perturbations, allowing the transformation to proceed through the direct formation and subsequent growth of the thermodynamically more stable state. As revealed by our previous \textit{ab initio} simulations in Pd-Ni-P systems~\cite{Shen2023Kinetics}, the coexistence transition is primarily driven by the motion of specific atoms (e.g., P), characterized by localized atomic rearrangements with low activation energy~\cite{Guan2012}. In contrast, when atomic mobility is sluggish, i.e., when the system resides in a deeper glassy state (e.g., at $T_{\mathrm{a}}=563$~K), the transition proceeds continuously. This process, however, involves a distinct structural rearrangement mechanism with a higher activation energy (\textbf{Fig.~6 and Fig.~S6}). This interpretation is also consistent with recent reports on high-entropy MGs, where suppressed atomic mobility promotes continuous GGT~\cite{Cao2024}.

The present results also suggest an intriguing analogy with molecular glass formers. In triphenyl phosphite (TPP), Tanaka and co-workers showed that nucleation-and-growth (NG)-type LLTs are typically accompanied the microcrystallite formation, whereas spinodal-decomposition (SD)-type LLTs are characterized by continuous fluctuations of structural order parameters~\cite{Tanaka2004,Kobayashi2016,Murata2015}. Although the characteristic small-angle scattering signatures reported for molecular systems~\cite{Murata2015} are not resolved in the present MG, additional enhanced scattering intensity near the FSDP is detected for the coexisting glass, even after short annealing times (\textbf{Fig.~5b}), whereas the continuous transition is accompanied by the gradual development of medium-range order (\textbf{Fig.~2}). These observations suggest that the coexistence and continuous transition mechanisms identified here may represent the MG counterparts of NG and SD-like mechanisms, respectively. Future experiments combining wide- and small-angle X-ray scattering will be important for testing this correspondence.

\begin{acknowledgments}

This work has received funding from the European Research Council (ERC) under the European Union's Horizon 2020 research and innovation programme (Grant Agreement No.~948780), the CNRS IRN project ``Spark'', the National Natural Science Foundation of China (Grant No.~92580120), the European Union--NextGenerationEU, Mission 4 Component 2 (CUP C93C24008240006), and the STARS@UNIPD program. IG and KO gratefully acknowledge the funding by the German Research Foundation through the grant GA 1721/4-2 on the “Thermodynamic analysis and modelling of Pd-Ni-P bulk metallic glass-forming system". We also
acknowledge the ESRF for
the provision of synchrotron radiation facilities under beamtime
proposal HC-5812 for the experiment and the long-term project
SC-5581 for commissioning the setup.

\end{acknowledgments}

\section*{DATA AVAILABILITY STATEMENT}
\vspace{-1.0\baselineskip}

The raw X-ray diffraction data are stored in the ESRF data repository at \url{https://doi.esrf.fr/10.15151/ESRF-ES-2007678307}. The raw data supporting the findings of this study are available from the corresponding authors upon reasonable request. 

\section*{AUTHOR CONTRIBUTIONS}
\vspace{-1.0\baselineskip}

JS, AC, VdL, GM, KO, BR, and MdM conducted the XRD--FDSC
experiments. JS performed the XRD and FDSC data analysis. PS and GM
designed, built, and commissioned the FDSC chamber. JS and BR wrote
the manuscript with contributions from all authors. BR and JS
conceived the study and led the investigation.

\bibliographystyle{apsrev4-2}
\bibliography{ref}

\end{document}